\pdfoutput=1

\documentclass[11pt,twoside,a4paper,note,final]{cms-tdr}
\def\svnVersion{a728d6a-D}\def\svnDate{2019/03/12}

\begin{document}

\hyphenation{had-ron-i-za-tion}
\hyphenation{cal-or-i-me-ter}
\hyphenation{de-vices}

\def\svnDate{2019/03/05}

\cmsNoteHeader{NOTE 2019-001} 
\title{Physics Potential of an Experiment using LHC Neutrinos}

\address[Hungary]{Hungarian Academy of Sciences, Inst. for Nuclear Research (ATOMKI), Debrecen, Hungary, and CERN,Geneva, Switzerland }
\address[cern]{CERN, CH-1211 Geneva 23, Switzerland}
\address[Napoli]{Universit\`a di Napoli Federico II and INFN, sezione di Napoli, Italy}
\address[Bologna]{INFN sezione di Bologna and Dipartimento di Fisica dell' Universit\`a, Bologna, Italy}
\address[NapoliC]{CERN, Geneva, Switzerland, and Universit\`a Federico II and INFN sezione di Napoli, Italy}
\address[Boston]{Boston University, Department of Physics, Boston, MA 02215, USA}
\address[ENEA]{INFN sezione di Bologna and ENEA Research Centre E. Clementel, Bologna, Italy}
\author[Hungary]{N. Beni}
\author[cern]{M. Brucoli}
\author[Napoli]{S. Buontempo}
\author[Bologna]{V. Cafaro}
\author[Bologna]{G.M. Dallavalle}
\author[cern]{S. Danzeca}
\author[NapoliC]{G. De Lellis}
\author[Napoli]{A. Di Crescenzo}
\author[Bologna]{V. Giordano}
\author[Bologna]{C. Guandalini}
\author[Boston]{D. Lazic}
\author[ENEA]{S. Lo Meo}
\author[Bologna]{F. L. Navarria}
\author[Hungary]{Z. Szillasi}

\abstract{
Production of neutrinos is abundant at LHC. 
Flavour composition and energy reach of the neutrino flux from proton-proton collisions depend on  the pseudorapidity $\eta$. 
At large $\eta$, energies can exceed the TeV, with a sizeable contribution of the $\tau$ flavour. 
A dedicated detector could intercept this intense neutrino flux in the forward direction, and measure the interaction cross section on nucleons in the unexplored energy range from a few hundred GeV to a few TeV.
The high energies of neutrinos  result in a larger $\nu$N interaction cross section, and the detector size can be relatively small. 
Machine backgrounds vary rapidly while moving along and away from the beam line.
Four locations were considered as hosts for a neutrino detector: 
the CMS quadruplet region (~25 m from CMS Interaction Point (IP)), UJ53 and UJ57 (90 and 120 m from CMS IP), RR53 and RR57 (240 m from CMS IP), TI18 (480 m from ATLAS IP).
The potential sites are studied on the basis of
(a) expectations  for neutrino interaction rates, flavour composition and energy spectrum,  (b) predicted backgrounds and in-situ measurements, performed with a nuclear emulsion detector and radiation monitors.
TI18 emerges as the most favourable location. 
A small detector in TI18 could measure, for the first time, the high-energy $\nu$N cross section, and separately  for $\tau$ neutrinos, with good precision,
already with 300 fb$^{-1}$ in the LHC Run3. 
}

\hypersetup{%
pdfauthor={CMS Collaboration},%
pdftitle={Physics Potential of an Experiment using LHC Neutrinos},%
pdfsubject={neutrino physics with LHC},%
pdfkeywords={CMS, physics, neutrino}}

\maketitle 

\section{Introduction}

Neutrinos in proton-proton (pp) interactions  at the CERN LHC arise promptly from W and Z leptonic decays,  b and c decays, and are produced in pion and kaon decays.
Their energies extend into the TeV range.
TeV-neutrino interactions on nucleons are currently uncharted physics
\cite{PDG}; 
only a few interactions of muonic neutrinos of cosmic origin with energies beyond 6 TeV were recorded in IceCube
\cite{IceCube}. 
LHC neutrinos offer the unique possibility  of probing the neutrino-nucleon interaction in the range from a few hundred GeV to a few TeV in laboratory.
Furthermore, 
the contribution of  the $\tau$ flavour to the LHC neutrino flux is sizeable.
$\tau$ neutrino interactions on nucleons are mostly unexplored, since so far only a handful of  events have been recorded 
(\cite{donut},
\cite{opera},
\cite{opera2})
at low energies.

The use of LHC as $\tau$ neutrino factory was first envisaged by De R\`ujula et al.  
\cite{derujula} and by Winter 
\cite{winter} 
and Vannucci
\cite{vannucci}.
in 1990-1993:  they  thought of a target detector  intercepting the very forward flux ( $\abs{\eta}>$7) of neutrinos (about 5\% have $\tau$ flavour) from b and c decays.
In a recent paper \cite{NUxshen} 
it was pointed out that  at larger angles (4$<\abs{\eta}<$5), leptonic W decays also provide an additional contribution 
to the neutrino flux ( about 30\% have $\tau$ flavour): in these events the charged lepton could be detected in coincidence
within the central $\eta$ region by existing experiments, ATLAS and CMS, providing an independent determination of the neutrino flavour. 

In ref.
\cite{NUxshen}
 it was proposed  to investigate the feasibility of a neutrino experiment at LHC, with the aim of:
\begin{itemize}
\item
measuring the cross section of high energy neutrino scattering on nucleon;
\item
collecting a sizeable sample of $\tau$ neutrino interactions.
\end{itemize}

Machine induced backgrounds constitute the major eperimental challenge. Simulations of LHC backgrounds 
were developed from the LHC design stage
\cite{Huhtinen} to
detailed predictions 
(\cite{Cerutti1},
\cite{Cerutti2},
\cite{Sophie})
validated with direct measurements.

In this paper, four locations are compared for hosting a neutrino detector: (i) a very near location, about 25 m from an Interaction Point (IP);
(ii) a  near location, about 90 and  120 m from IP;
(iii) a far location, about 240 m from IP;
(iv) a very far location, about 480 m from the IP.
For each location, 
expectations  for neutrino flux, interaction rate, flavour composition and energy spectrum are calculated, and the LHC background
is investigated with simulations, 
and 
in-situ measurements performed with nuclear emulsions and radiation monitors.

The paper is structured as follows. Section 2 describes the general properties of neutrinos from pp collisions at LHC.
Sections 3 to 6, each dedicated to a location, discuss site suitability for hosting a detector and investigate the physics potential in terms of neutrino interaction rate and background environment for an integrated luminosity of 3000 fb$^{-1}$ (HL-LHC).
Section 7 draws the conclusions and hints at an experiment feasible already with 300 fb$^{-1}$ for the LHC Run3 in 2021-2023.

\section{Characteristics of LHC Neutrinos}

\begin{figure} [h]
\centering
\includegraphics[width=0.7\textwidth]{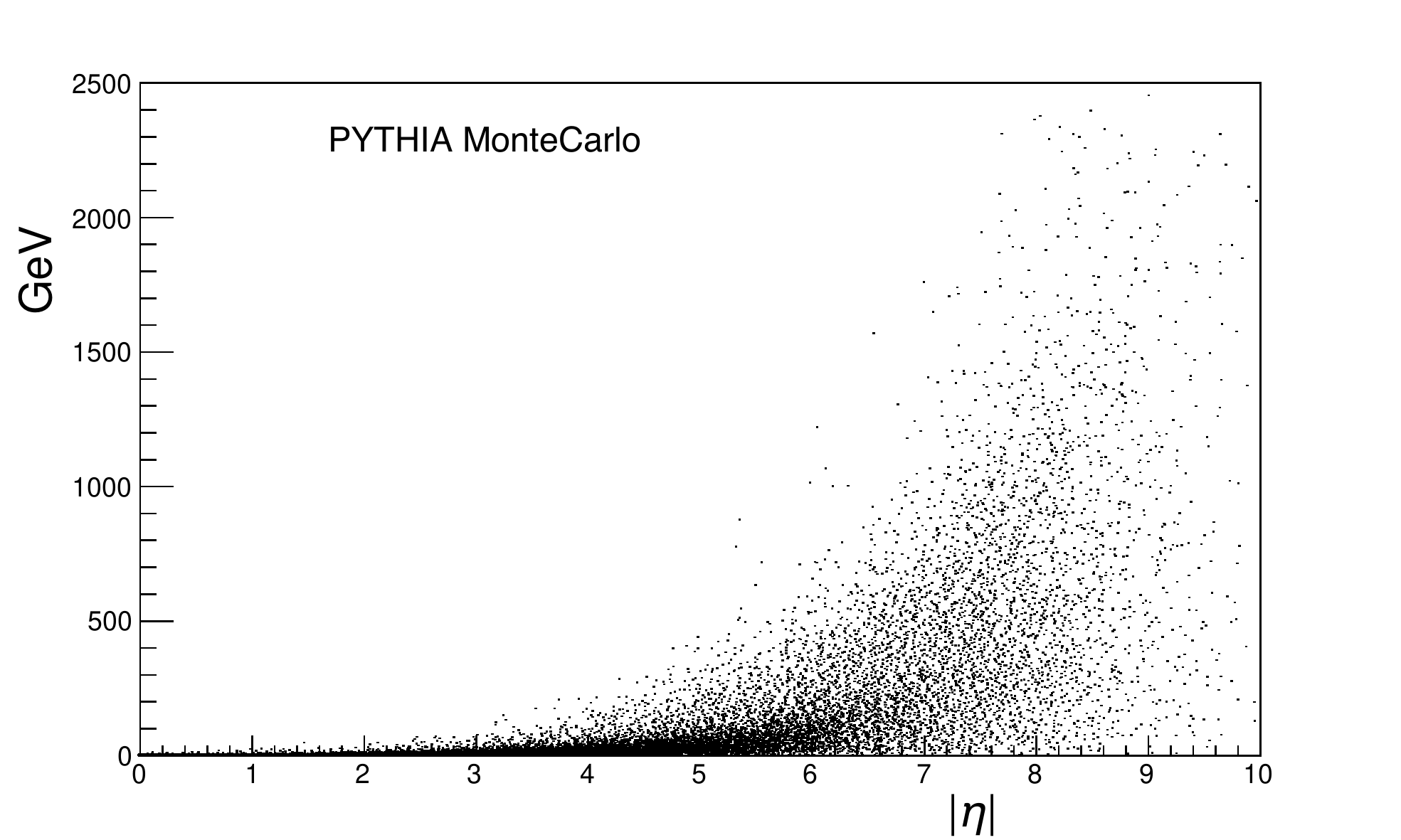}
\caption[Caption scatter plot bc] {Scatter plot of neutrino energy versus pseudorapidity $\eta$ in b and c decays. \label{fig:scatterbc}}
\end{figure}

\begin{figure} [h]
\centering
\includegraphics[width=0.7\textwidth]{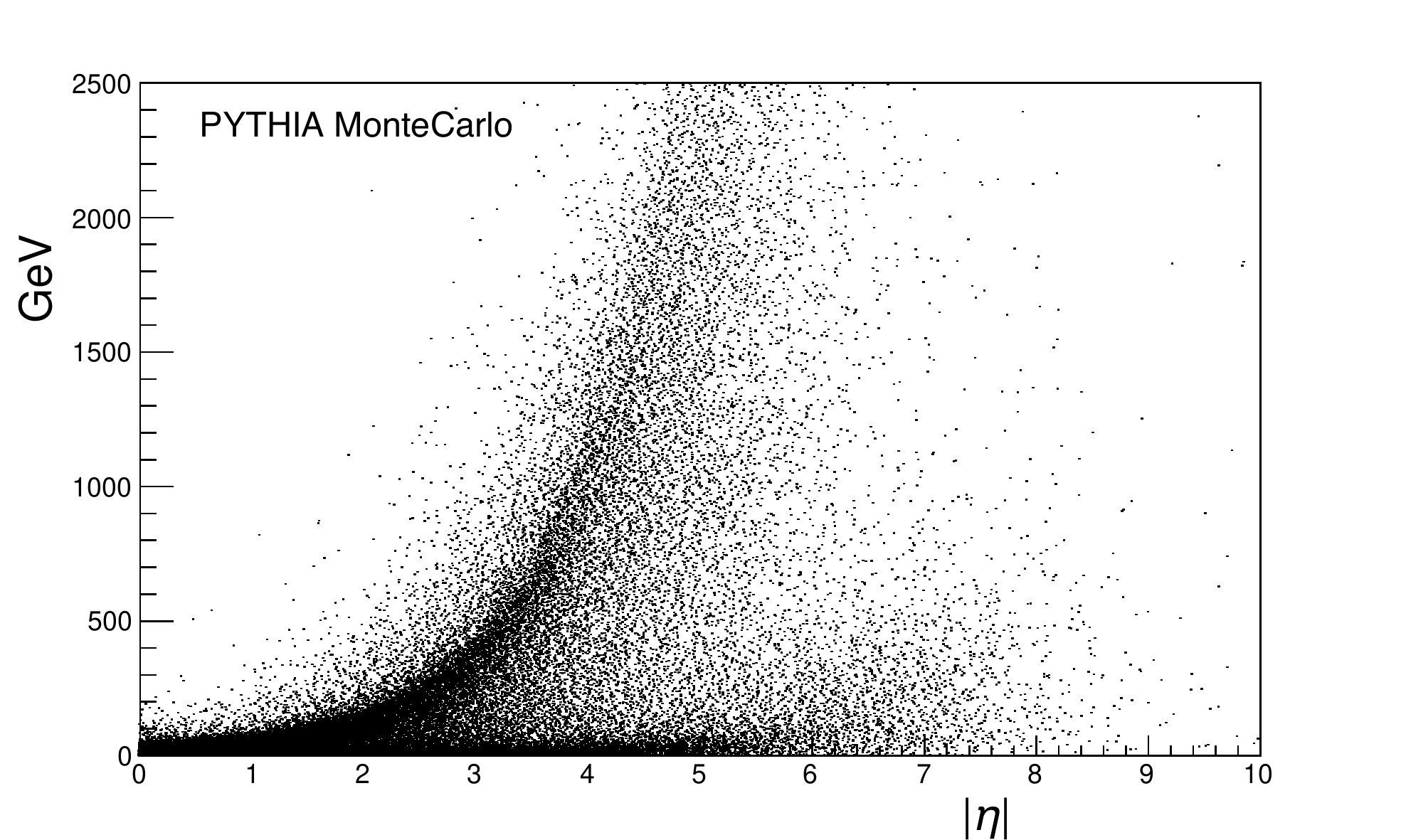}
\caption[Caption scatter plot W] {Scatter plot of neutrino energy versus pseudorapidity $\eta$ in pp events with W production. Neutrinos from the leptonic W decays are seen to be kinematically well separated. \label{fig:scatterW}}
\end{figure}

Proton-proton (pp) collision events at $\sqrt{s}=$ 14 TeV were simulated using Pythia 8.226
\cite{Pythia}. 
Figures 
~\ref{fig:scatterbc} and
~\ref{fig:scatterW} 
show scatter plots of energy versus pseudo-rapidity \abs{\eta} separately for neutrinos in pp events with b and c production 
and in pp events with W production.
Neutrinos from heavy quarks are abundant, and are very forward;
$\tau$ neutrinos account for about 5\% of the neutrino flux. 
Because of the Lorentz boost along the beam axis, 
their energies can  attain the TeV range when
$\abs{\eta}$ is 6.5 or larger.   
In the region of  3$<\abs{\eta}<$6 prompt neutrinos from leptonic W decays, and similarly from Z decays, give
a consistent contribution to the TeV neutrino tail;  
about 30\% of these are $\tau$ neutrinos.


The W is boosted along the beam line, so that the charged lepton and the neutrino from the decay tend to be on the same side in pseudo-rapidity, and they are back-to-back in the plane transverse to the beam. When the neutrino lies in  4$<\abs{\eta}<$5, the charged lepton is within  $\abs{\eta}<$3 in about 50\% of the cases, i.e. it can be observed in the ATLAS or CMS detectors, providing an independent tagging of the neutrino species.

\begin{figure} [h]
\centering
\includegraphics[width=0.7\textwidth]{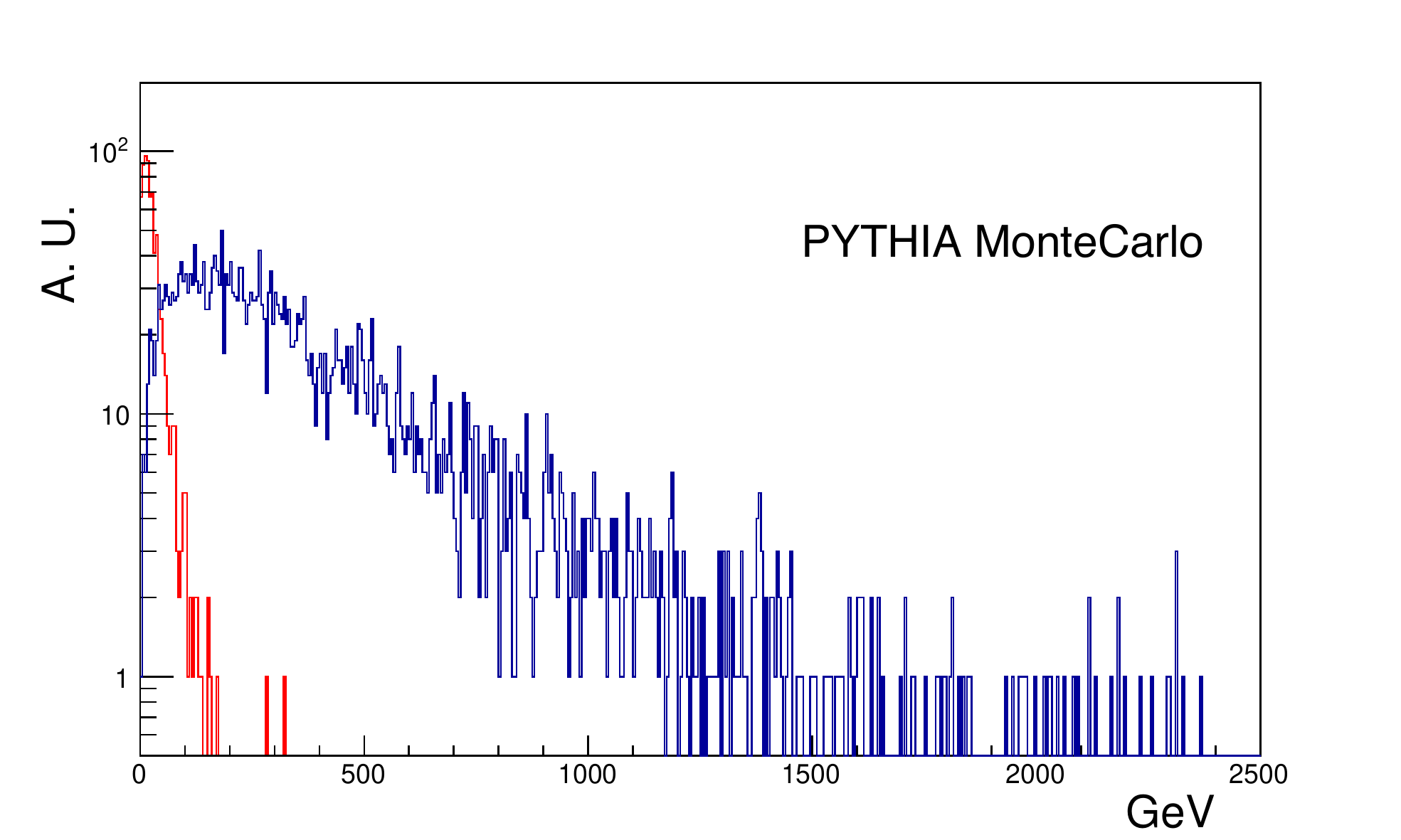}
\caption[Caption scatter plot bc] {Energy spectra for neutrinos from b and c when neutrino has 4$<\abs{\eta}<$5 (lred) and when $\abs{\eta}>$ 6.5 (blue). \label{fig:spectrabc}}
\end{figure}

\begin{figure} [h]
\centering
\includegraphics[width=0.7\textwidth]{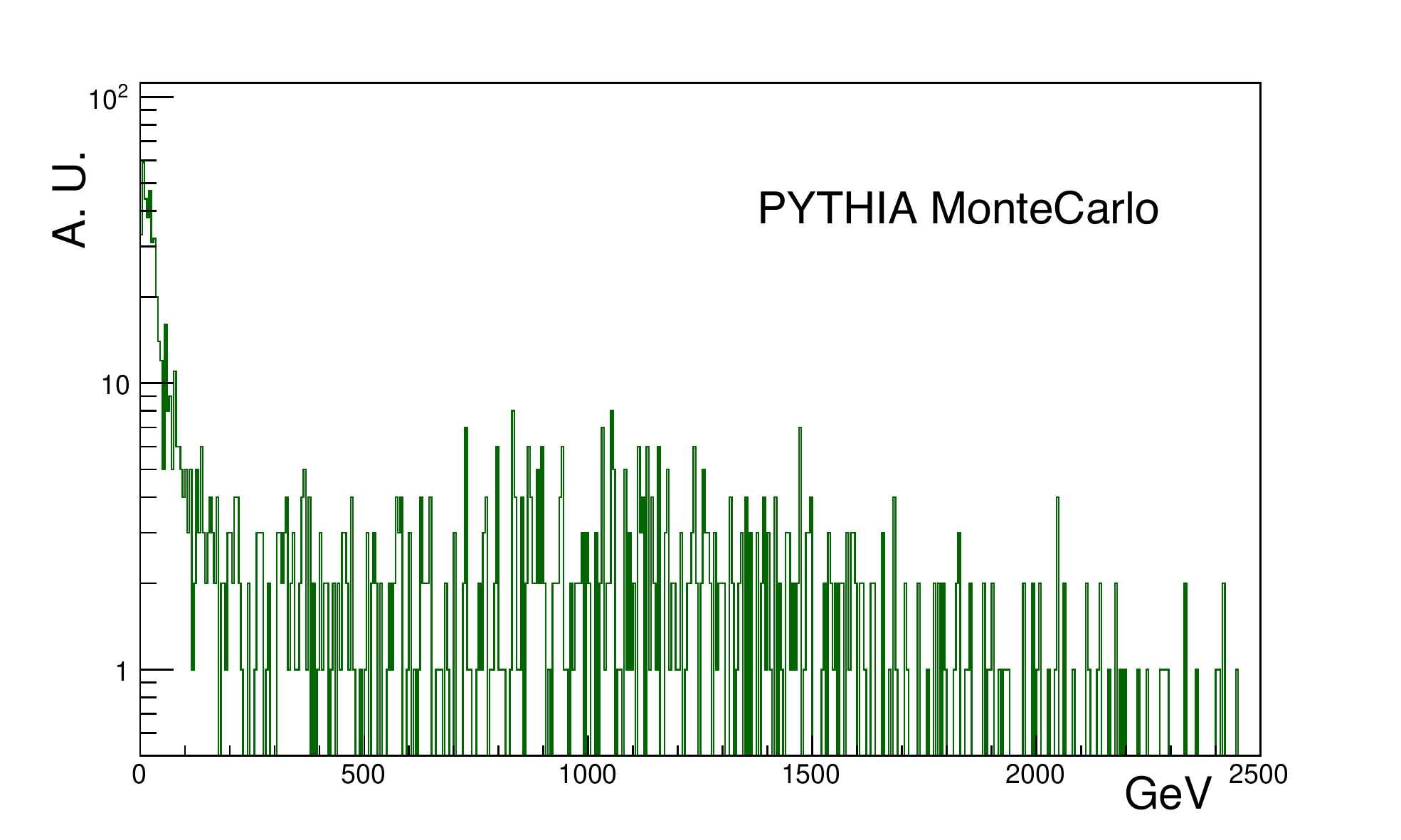}
\caption[Caption scatter plot W] {Energy spectrum for neutrinos from W when neutrinos are in 4$<\abs{\eta}<$5. \label{fig:spectraW} }
\end{figure}

Figures
~\ref{fig:spectrabc} and
~\ref{fig:spectraW}  
show the neutrino energy spectra for two slices in $\abs{\eta}$. 
Figure 
~\ref{fig:spectrabc}
is for neutrinos from b and c decay: in 4$<\abs{\eta}<$5 the spectrum is confined to low energies while in $\abs{\eta}>$ 6.5 the spectrum reaches very high energies. Although the absolute scale is arbitrary, the relative normalization is based on the differential cross section and is correct.
The spectrum of neutrinos from Ws when 4$<\abs{\eta}<$5 is plotted in Fig. 
~\ref{fig:spectraW} showing the high energy contribution of leptonic W decays. 


In the remainer of this paper, we refer to W or Z production as W production. The contribution to the neutrino flux from Z is about 20\% of that from W.
Table ~\ref{tab:fluences}  compares neutrino fluences for 3000 fb$^{-1}$ of pp collision events  in two regions  4$<\abs{\eta}<$5, $\abs{\eta}>$ 6.5, separately for heavy quark production and for W production. For b and c production,  although the fluence integrated in the range 4$<\abs{\eta}<$5 is slightly larger, the neutrino energy there is lower, therefore the $\nu$N cross section will be smaller. For W production, 
although the total flux is smaller than for b and c decays, 
the flux has a large component at high energies.

\begin{table} [h]
\begin{center}
\topcaption{ Neutrino fluences in 4$<\abs{\eta}<$5, $<\abs{\eta}>$ 6.5, for 3000 fb$^{-1}$. }
\label{tab:fluences}
\begin{tabular} {lcc} \hline
 & 4$<\abs{\eta}<$5 & $\abs{\eta}>$6.5 \\ \hline
Neutrinos from pp collisions at $\sqrt{s}$=14 TeV with c and b production &  & \\ \hline
neutrino fluence in acceptance /3000 fb$^{-1}$ & 3.9 10$^{14}$ & 1.7 10$^{14}$ \\
$\tau$ neutrino fluence in acceptance /3000 fb$^{-1}$ & 2.7 10$^{13}$ & 9.3 10$^{12}$ \\
neutrino average energy (RMS) (GeV) & 30(30) & 400(350) \\ \hline \hline
Neutrinos from pp collisions at $\sqrt{s}$=14 TeV  with W production  &  & \\ \hline
neutrino fluence in acceptance /3000 fb$^{-1}$ & 1.7 10$^{10}$ & 1.1 10$^{9}$ \\
$\tau$ neutrino fluence in acceptance /3000 fb$^{-1}$ & 4.7 10$^{9}$ & 1.1 10$^{8}$ \\
neutrino average energy (RMS) (GeV) & 600(600) & 400(400) \\  \hline
\end{tabular}
\end{center}
\end{table}

In view of a possible experiment using LHC neutrinos it should also be noted that
the $\nu$N cross section grows with energy: low energy experiments need a large target mass; at energies about the TeV the detector could be very compact.

\section{Very Near (VN) Location}

\subsection{Geometrical Acceptance}

Near the CMS  IP,  the area surrounding the focusing magnet (Q1) closest to CMS in the LHC tunnel matches the
4$<\abs{\eta}<$5 acceptance.  
The CMS forward hadronic calorimeter (HF) --- 10 nuclear interaction lengths --- and the CMS forward shield ---  composed of steel and borated concrete, with a total of 26 nuclear interaction lengths --- provide an excellent shield against all particles emerging from the IP, except for muons and neutrinos.
Although the triplet of final focus magnets on both sides of the IP will be rebuilt for HL-LHC, the current geometry was assumed for calculating an approximate rate of neutrino interactions.     
A hypothetical cylindrical detector placed at 25 m from the IP, around the Q1 magnet, with inner and outer radii of  35 and 90 cm, could  cover the  4$<\abs{\eta}<$5 range in most of the azimuthal acceptance (320 $\deg$ of 360) and be stretched in  length for over 7 meters, up to about 20. 
The detector is placed at one side of the IP only, assume the positive z end 4$<\eta<$5.
The target mass in the detector is taken  to be 
4.1 $\times$ 10$^{27}$
 nucleons/cm$^2$, corresponding to about 6 m of lead over 15 m, for a weight of about 10 ton/m.
The LHC beam line  elements limit extensions of  the target length that may be needed  for increasing the detector mass and the $\nu$ interaction rate. 

\subsection{Neutrino Interaction Rate}

In the calculation of the neutrino interaction rate, an event by event weight is applied for properly taking into account 
the $\nu$N interaction cross-section dependence on energy and on neutrino type
\cite{PDG}.
For the Charged Current cross section of $\tau$ neutrinos on nucleons, we can only refer to theoretical expectations 
\cite{nutauN},
that take into account the effect of the F4 and F5 structure functions
\cite{F4F5}.
As an example, the interaction of a muon neutrino of 30 GeV on nucleons has a cross section of  about 20 $\times$10$^{-38}$ cm$^{2}$, and 
 a $\tau$ neutrino of 600 GeV  of about 350 $\times$10$^{-38}$ cm$^{2}$;  for anti-$\nu$ the cross section is about a half that of $\nu$.
Table ~\ref{tab:verynear}  summarizes the expected neutrino interactions for 3000 fb$^{-1}$ of pp collision events, separately for heavy quark production and for W production. 

\begin{table} [h]
\begin{center}
\topcaption{ Expected neutrino interactions in the VN (about Q1) location, for 3000 fb$^{-1}$. }
\label{tab:verynear}
\begin{tabular} {lr} \hline 
& 4$<\eta<$5 , +z end \\ \hline
Neutrinos from c and b production in pp collisions at $\sqrt{s}$=14 TeV  &  \\ \hline
total neutrino interactions & 2.7$\times$ 10$^{5}$ \\
neutrino interactions with E$>$100 GeV & 46000\\ 
neutrino interactions with E$>$300 GeV & 5500\\ \hline
$\tau$ neutrino interactions & 8500 \\ \hline \hline
Neutrinos from W production in pp collisions at $\sqrt{s}$=14 TeV  &  \\ \hline
total neutrino interactions & 320 \\
neutrino interactions with E$>$100 GeV & 310\\ 
neutrino interactions with E$>$300 GeV & 300\\ \hline 
$\tau$ neutrino interactions & 110\\
\end{tabular}
\end{center}
\end{table}

The bulk of the event sample is from b and c decays, with an energy spectrum quickly decreasing, while leptonic Ws give a small contribution, 
which is concentrated at very high energies and becomes dominant for energies above 500 GeV.
With respect to the experiment targets:
\begin{itemize}
\item{\rm high energy $\nu$N cross secion: }
a measurement of the neutrino interaction cross section on nucleons at  energy above  300 GeV can be perfomed to a  few percent accuracy.
Thanks to the leptonic W contribution, the accuracy will remain better than 10\% at E $=$ 1 TeV. In a sample of a hundred neutrino interactions, the charged lepton from leptonic W decays will be observed in CMS.
\item{\rm $\tau$ neutrino sample: }
a few thousand $\tau$ neutrinos can be observed. A hundred  will have energies in the TeV range, thanks to the W contribution.
\end{itemize}

Hence, a detector placed in this location  has an interesting physics potential. A drawback is its interference with the design of the focusing magnets for HL-LHC.

\subsection{Backgrounds, Radiation Environment}

Measurements by LHCb
\cite{LHCb_charged}
 and TOTEM 
\cite{TOTEM_charged}
show that in 4$<\abs{\eta}<$5 there are about 4.5 charged particles on average per pp inelastic interaction at $\sqrt{s}$ = 13 TeV, mostly pions and kaons. With a pp inelastic cross section of 80 mb and a luminosity of 2 $\times$ 10$^{34}$ cm$^{-2}$ s$^{-1}$, it is estimated that in 4$<\abs{\eta}<$5   there are 7$\times$ 10$^{9}$ charged particles per second. 
However, in the location under study  the detector  would be shielded from charged hadrons emerging from the IP  by the CMS hadronic calorimeter (HF) and the CMS forward shield, providing together 36 nuclear interaction lengths, i.e. a damping factor of 2.3x10$^{-16}$, corresponding to almost full absorption for hadrons.
Muons from pion and kaon decays will be rare: $\pi$ and $K$ particles with energies larger than a few GeV will interact before they can decay, and, if not, only energetic muons from the decays will penetrate the forward shield. 

The muon rate from pp collisions at $\sqrt{s}$ = 13 TeV in 4$<\abs{\eta}<$5 was measured by LHCb
\cite{LHCb_muon}: 
in minimum bias events they observe a rate of 25 kHz of muons with p$>$10 GeV at a luminosity of 4 $\times$ 10$^{32}$ 
cm$^{-2}$ s$^{-1}$.  This is in excellent agreement with Pythia calculations, assuming that the dominant source is b and c decays:  the production cross section is  48$\times$10$^5$  nb at  $\sqrt{s}$ = 13 TeV and 1.4\% of these events have a muon in 4$<\abs{\eta}<$5, which gives an expectation of  27 kHz. The predicted muon rate at 2 $\times$ 10$^{34}$ cm$^{-2}$ s$^{-1}$ luminosity  is 1.25 MHz, however, when considering the muon energy spectrum only $<$10\% of those muons are expected to reach the detector in Q1, due to the CMS magnetic field and shielding material.  
In the proposed location, the detector has a cross section of about 2.2 x 10$^4$ cm$^2$, so that about 6 muons per cm$^2$ will impinge on it in a second, i.e. a few $\times$10$^5$ /cm$^2$/fb$^{-1}$.

Machine induced backgrounds vary rapidly while moving along and away from the beam line. An occupancy of a few per cm$^2$ per second is estimated for muons in the beam halo moving towards CMS
\cite{Huhtinen}. 
A sizeable contribution might come from hadronic secondaries emerging from the pp collisions at lower angles and interacting with the machine elements, like the TAS collimators positioned at 18 m from the CMS IP
\cite{Sophie}: 
the energy and angular distributions of the debris particles have large uncertainties, but the proposed location should lie in a local minimum.
The flux of neutrons and photons in early estimations ranges from 10$^{-3}$ to 1 /cm$^2$/s
\cite{Huhtinen}. 
Later simulations  
\cite{Cerutti1, Cerutti2}, 
validated with local measurements, show a very large contribution of thermal neutrons. Their flux has local
fluctuations
by several orders of magnitude along the beam line, and can reach 10$^4$ /cm$^2$/s, i.e about 10$^9$ /cm$^2$/fb$^{-1}$ ; 
the uncertainty in the proposed location is very large.

\subsubsection{In-situ measurement}

A measurement was performed to directly evaluate the background environment.
A stack of emulsions  was installed  in the proposed location at the beginning of the 2018 LHC tun
\cite{Laza}.  
It was placed on the concrete underneath Q1, at about 25 m from the CMS IP, covering the region of 80 to 90 cm from the beam axis, i.e. at pseudorapidity $\eta$ about 4.
It was retracted after CMS had collected 1.6 fb$^{-1}$ of luminosity.

\begin{figure} [b]
\centering
\includegraphics[width=0.7\textwidth]{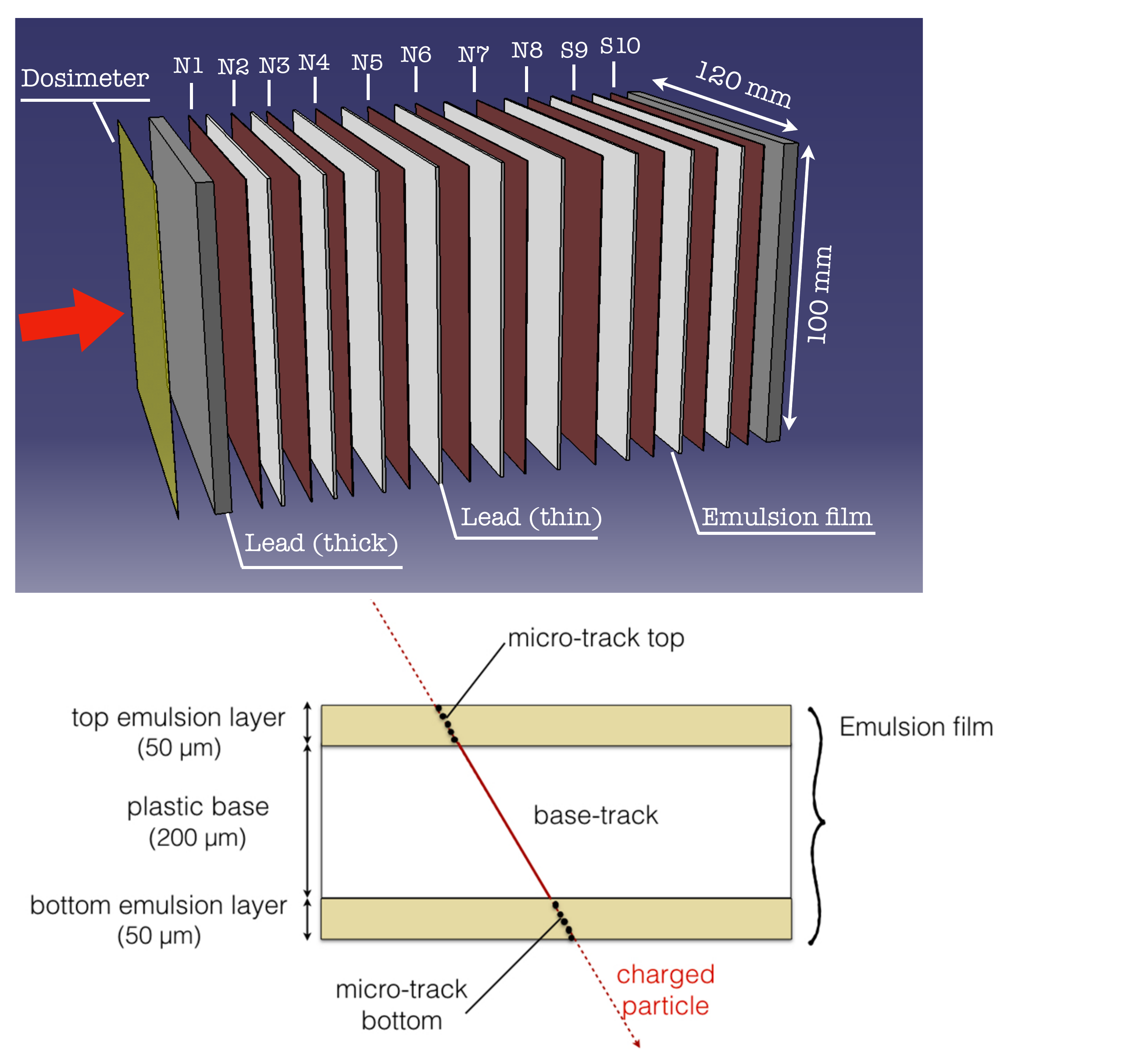}
\caption[Caption emulsions] {Structure of an emulsion stack used for the measurements. Also shown a sketch of a charged particle crossing an emulsion film. \label{fig:emulsions}}
\end{figure}

The stack consisted of 10 layers of nuclear emulsions interleaved with 1 mm lead sheets (Fig.
~\ref{fig:emulsions}).
The package was protected from neutrons and photons by a 30 mm thick layer of borated polyethylene and 5 mm of lead,
on all sides but at the bottom.
A 5 mm air gap was left at the bottom so that the inside would not heat up and damage the emulsions.  
Polyethylene doped with 5\% $B_{2}O_{3}$ has an attenuation length for thermal neutrons of 2.4 mm.
Simulations with Geant4 v10.4
\cite{geant} 
were performed to check the attenuation lenght of borated polyethylene for neutrons 
with energy from 0.025 eV to 1.0 GeV and to study the effect of the thin air gap at the bottom, 
since neutrons (mainly with thermal energy)  come from all directions and can flow through the thin gap; the effective damping factor provided by the
$B_{2}O_{3}$ layer was estimated to be  10$^{-3}$. 
The Physics List (PL) used was 
FTFP INCLXX HP that relies on the Fritiof + pre-equilibrium
model (FTFP) at high energy, the INCL++
\cite{LoMeo1}
model at intermediate energies, and the NeutronHP model for a detailed treatment of neutrons below 20 MeV. This PL has been successfully benchmarked
against experimental results in Ref.
\cite{LoMeo2}.

The package became radioactive during the exposure. An activity of  35 $\mu$Sv/h was measured immediately after extraction, 
with a long decay time.
When the activity fell below safety threshold, the gamma spectrum from the emulsions was measured. It showed lines charactestics of the decay of Ag$^{110m}$, a metastable state produced when a neutron is captured by a Ag$^{109}$ atom.
The measurement was repeated after a month. From the intensity of the lines, 
the amount of Ag$^{110m}$ could be determined, which, knowing the emulsion composition and  
the neutron-silver cross section
\cite{EXFOR}, 
can be turned into a measurement of the integrated neutron flux.
However the neutron flux was delivered over an extended time, following the LHC running schedule.
Thus the production rate of Ag$^{110m}$ had to be weighted over time using the luminosity information provided by CMS,
and the integrated fluence of neutrons passing the borated polyethylene layer and impinging on the emulsions was determined to be of the order of 10$^8$ /cm$^2$/fb$^{-1}$, comparable to the expectations with no $B_{2}O_{3}$ protection. 

The emulsions finally underwent photographic development. The hit density was far above the tolerable limit 
--- about 10$^{7}$ /cm$^{2}$ (see section 5.3.1 ) --- 
for proceeding to further analysis.

The observed neutron flux, larger than the most conservative expectations, and  its predicted large variability along the beam line, discourage the use of this location.

\section{Near (N) Location}

\subsection{Geometrical Acceptance}
The UJ57  and UJ53 halls lie at 90 and 120 meters, respectively, from the CMS IP, where the beam line meets the ends of the tunnel that bypasses the CMS experimental cavern, on oppostite sides of CMS.
They are open caverns --- long and narrow --- on a side of the beam line.
An iron wall could be built for providing isolation from the beam line and lowering the radiation level if necessary. 
The available volume for installing a detector is limited, since a 5 m clearance from the beam line is reserved 
for transport of accelerator elements.
It could be 2 m wide, 2 m high, and extend for 15 m. Given the large size iron is preferred to lead. 
The target mass could consist of 9 meters of iron in total, weighting 20 ton/m and
providing 4.0$\times$10$^{27}$ nucleons/cm$^2$.
The detector would lie at 5 m from the beam, giving a reduced azimuthal coverage (about 1/18 of 2$\pi$). The detector in UJ57 can cover 
the pseudorapidity range -3.6$<\eta<$-3.2, while the one in UJ53 can cover 3.5$<\eta<$3.9.

\subsection{Neutrino Interaction Rate}

From the studies  in Section 2, the $\eta$ acceptance in this location is not optimal. The energy spectrum of neutrinos from b and c decays 
as well as for leptonic W decays will be softer,
the tail at very high neutrino energies will be  limited marginally populating the 1 TeV region and beyond. 
The kinematical acceptance for leptonic W decays with the charged lepton in the CMS acceptance will be enhanced, but that advantage vanishes because of the small angular coverage in azimuth. 

It is assumed that both UJ53 and UJ57 woul be equipped. Table ~\ref{tab:far}  summarizes the expected neutrino interactions for 3000 fb$^{-1}$ of pp collision events, separately for heavy quark production and for W production. 

\begin{table} [h]
\begin{center}
\topcaption{ Expected neutrino interactions in the near (N) (UJ53+UJ57) location, for 3000 fb$^{-1}$. }
\label{tab:far}
\begin{tabular} {lr} \hline 
& 3.2$<\eta<$3.9 \\ \hline
Neutrinos from c and b production in pp collisions at $\sqrt{s}$=14 TeV  &  \\ \hline
total neutrino interactions & 2.0$\times$ 10$^{3}$ \\
neutrino interactions with E$>$100 GeV & 110\\ 
neutrino interactions with E$>$300 GeV & 1\\ \hline
$\tau$ neutrino interactions & 240 \\ \hline \hline
Neutrinos from W production in pp collisions at $\sqrt{s}$=14 TeV  &  \\ \hline
total neutrino interactions & 13 \\
neutrino interactions with E$>$100 GeV & 12\\ 
neutrino interactions with E$>$300 GeV & 10\\ \hline 
$\tau$ neutrino interactions & 4\\
\end{tabular}
\end{center}
\end{table}

With respect to the experiment targets:
\begin{itemize}
\item{\rm high energy $\nu$N cross-secion: }
in this location, only a very coarse measurement of
the neutrino interaction cross section on nucleon at  energy above  300 GeV can be performed, 
relying on a few  leptonic W decays, with accuracy not better than 40\%;  it is impracticable at 1 TeV.
\item{\rm $\tau$ neutrino sample: }
a few hundred $\tau$ neutrinos can be observed. Their energies will be low, averaging to about 15 GeV.
\end{itemize}

Hence, a detector placed in this location  has a marginal potential for neutrino physics.

\subsection{Backgrounds, Radiation Environment}

The UJ57 hall is located at the end of the magnet triplet region, of which Q1 (VN location) is  the closest to CMS. 
In reference
\cite{Cerutti2}, 
expected machine induced backgrounds in UJ57 have been calculated, and found of similar composition and level as in 
the VN location.
The charged hadron fluence will exceed 10$^6$/cm$^2$/fb$^{-1}$
and that of thermal neutrons will be larger than  10$^8$/cm$^2$/fb$^{-1}$.
The situation is predicted to improve in the shielded areas farther downstream the beamline, discussed in section 5.
The flux of muons, at 90 m from the IP, although it is distributed over a large area,  is intense, $\sim$3$\times$10$^5$/cm$^2$/fb$^{-1}$.

Given the low potential that this location has shown for physics, no  in-situ measurement of the background was planned for confirming the calculations.

\section{Far (F) Location}

\subsection{Geometrical Acceptance}

Further away from the IP along the beam line, on either sides of CMS, are situated the RR53 and RR57 halls, that are classified as LHC "alcoves". 
Those are radiation shielded areas that can host electronic equipment made of conventional not radiation tolerant electronics. 
The radiation environment in those halls is intrinsically mild by construction.
The RR53 and RR57 alcoves are located at 237 meters from the CMS IP, at a distance of 3 meters from the beam line, and shielded from it by a 40 cm thick iron wall. The hall extends for 20 meters parallel to the beam, and has transverse dimensions of over 6 meters horizontally and 4 meters vertically. The iron wall is discontinued at both extremes of the hall for giving access.
Slightly upstream of RR53 (RR57) along the beam line are the TOTEM roman pots and the TCL6 collimator.

A detector occupying the RR53 (RR57) hall would subtend  the pseudirapidity region 4$<\abs{\eta}<$5, the same as for
the detector in the VN location (section 3). 
Its transverse dimensions would need to be large, 5m horizontally and 3 m vertically.
The length is limited to at most 20 m by the hall size.
Given the large transverse dimensions, iron is used as target.
The target detector could feature
3.0 $\times$ 10$^{27}$
 nucleons/cm$^2$, corresponding to 6.5 m of iron, for a weight of about 40 ton/m.

Although at a distance of 240 m from the IP, a coincidence with a charged lepton from a leptonic W decay in CMS could still be envisaged. 
Proof is that the nearby TOTEM roman pots can be operated together with CMS
\cite{TOTEM_trigger}.

\subsection{Neutrino Interaction Rate}

Event kinematics is as shown for the VN location, because it covers the same 4$<\abs{\eta}<$5 range. 
However, since the detector lies at a distance by the side of the beam line,  the azimuthal acceptance is reduced and that cuts down  the event statistics. 
An azimutal acceptance of 30 degrees (8\% of 2$\pi$) can be achieved.
The acceptance loss can be partially recovered by equipping both RR53 and RR57.

Table ~\ref{tab:far}  summarizes the expected neutrino interactions for 3000 fb$^{-1}$ of pp collision events, separately for heavy quark production and for W production. 

\begin{table} [h]
\begin{center}
\topcaption{ Expected neutrino interactions in the far (F) (RR53) location, for 3000 fb$^{-1}$. }
\label{tab:far}
\begin{tabular} {lr} \hline 
& 4$<\eta<$5 , +z end \\ \hline
Neutrinos from c and b production in pp collisions at $\sqrt{s}$=14 TeV  &  \\ \hline
total neutrino interactions & 2.5$\times$ 10$^{4}$ \\
neutrino interactions with E$>$100 GeV & 4300\\ 
neutrino interactions with E$>$300 GeV & 520\\ \hline
$\tau$ neutrino interactions & 880 \\ \hline \hline
Neutrinos from W production in pp collisions at $\sqrt{s}$=14 TeV  &  \\ \hline
total neutrino interactions & 30 \\
neutrino interactions with E$>$100 GeV & 29\\ 
neutrino interactions with E$>$300 GeV & 28\\ \hline 
$\tau$ neutrino interactions & 10\\
\end{tabular}
\end{center}
\end{table}

The bulk of the event sample is from b and c decays, with a fast decreasing energy spectrum.
Leptonic W decays give a small contribution concentrated at very high energies, that becomes dominant for energies larger than 500 GeV.
With respect to the experiment targets:
\begin{itemize}
\item{\rm high energy $\nu$N cross-secion: }
a measurement of the neutrino interaction cross section on nucleon at  energy $>$  300 GeV can be perfomed to a  better than 10\% accuracy, and to 
20\% when E is 1 TeV. In a few events, the charged lepton from leptonic W decays will be observed in CMS.
\item{\rm $\tau$ neutrino sample: }
nearly a thousand $\tau$ neutrinos can be observed. A few  will have TeV energies, thanks to the W contribution.
\end{itemize}

Hence, a detector placed in this location  has physics potential, but the sample of detected neutrinos from leptonic W decays with the charged lepton observed in CMS  woul be marginal.  
Another drawback is the size of the detector, which implies a big investment.

\subsection{Backgrounds, Radiation Environment}

Regarding the charged particle flux fom the IP, the arguments of section 3.3 apply to this location as well. The flux is still dominated by muons, which distribute over a larger area, resulting in at most  10$^5$ /cm$^2$/fb$^{-1}$.

Machine induced backgrounds in RR53 (and similary in RR57) were simulated in detail 
and compared to measurements taken with Radiation Monitor devices 
over long LHC running periods
\cite{Cerutti2} .
The radiation field inside the hall, although generally low,  is neither uniform in space nor in time.
The main component is hadronic secondaries with energies larger than a MeV, that spray out from elements along the beam line. 
The radiation background is more intense near the entrances where the iron wall is discontinued, 
it can reach 10$^7$-10$^8$ hadrons /cm$^2$/fb$^{-1}$), and its intensity depends on the operation mode of the machine, in particular the TCL6 collimator, located about 20 m upstream the hall. 
Away from the entrance, at about 6 meters from the beam line and in the center of the hall, the fluence of hadronic secondaries is predicted to be mild (about 10$^{4}$-10$^{6}$/cm$^{2}$/fb$^{-1}$);  the intensity increases  by an order of magnitude  at the side closest to the beam line,  and drops  as much at the opposite side of the hall. 
Muons can emerge from the decay of hadronic secondaries, and their flux is estimated in the order of 10$^{2}$/cm$^{2}$/fb$^{-1}$) in the center of the hall.
The neutron background is dominated by thermal neutrons: their fluence could reach up 10$^8$-10$^9$ /cm$^2$/fb$^{-1}$ in the hottest area. 

\subsubsection{In-situ  measurement}

An emulsion and lead stack similar to the one used in section 3.3.1 was prepared. 
It was installed in RR53 during the LHC Technical Stop 2, in September 2018, and retracted after 5.4 fb$^{-1}$ of pp collisions collected by CMS.

\begin{figure} [h]
\centering
\includegraphics[width=1.0\textwidth]{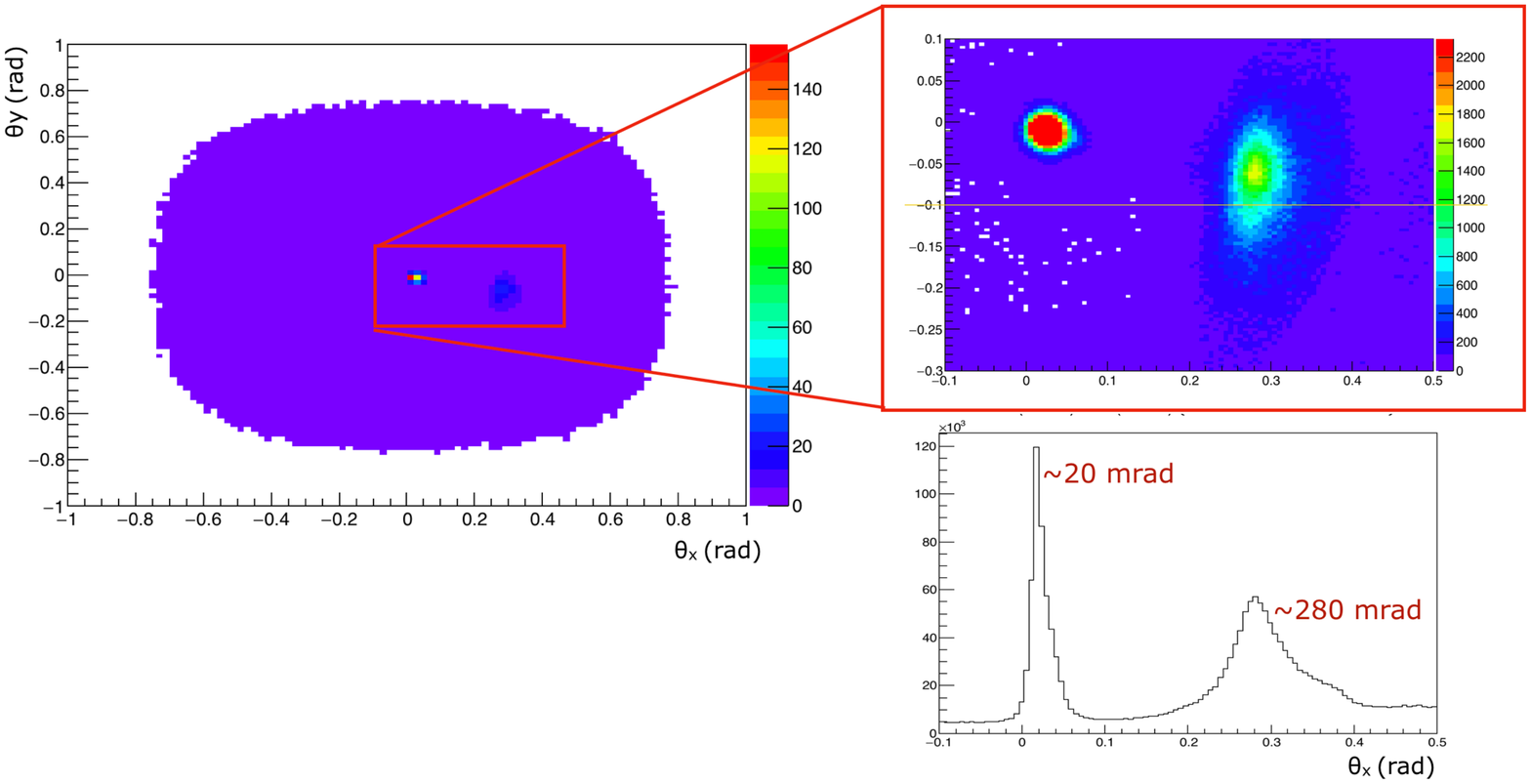}
\caption[Caption scatter plot ADC] {Distributions of the measured inclinations of the observed particle tracks, horizontally ($\theta_{x}$) and vertically, ($\theta_{y}$) in one of the emulsion films of the package. \label{fig:ADC_plot}}
\end{figure}

The thickness of the borated polyethylene container was increased to 90 mm, providing an attenuation factor close to 10$^{-16}$ for thermal neutrons, but it was open at the bottom of the box in order to prevent overheating. 
The package was placed in the approximate center of the RR53 hall, 5.5 m aside of the beam line, and 1 m below it, on a grating false floor to diperse heat from the box.
The CERN Engineering team provided Radiation Monitor devices 
\cite{radmon1, radmon2}
that were placed in critical positions in the hall in order to map the radiation field both inside and in proximity of the hall.
In each position two devices were installed, with different thresholds, so to infer both 
the  high energy hadron and the thermal neutron components separately.
The monitors were retracted at the same time of the emusion package, and analysed 
\cite{Danzeca}.
Those near the center of the RR53 hall, close to the emulsions, measured mild fluences of both high energy hadrons and of thermal neutrons  
of a few $\times$ 10$^6$ /cm$^2$ / fb$^{-1}$. 
The meaured values  increased closer to the iron wall, or  near the forward and backward entrances, by almost an order of magnitude. 
For comparison, two measurements were performed during the same period outside the hall, 
along the beam line, and showed values two orders of magnidute larger. 
These measurements are in average in agreement with the predictions 
\cite{Cerutti2} for the LHC machine running configuration used during the time of the measurements, and provide more detailed information.

\begin{figure} [t]
\centering
\includegraphics[width=1.0\textwidth]{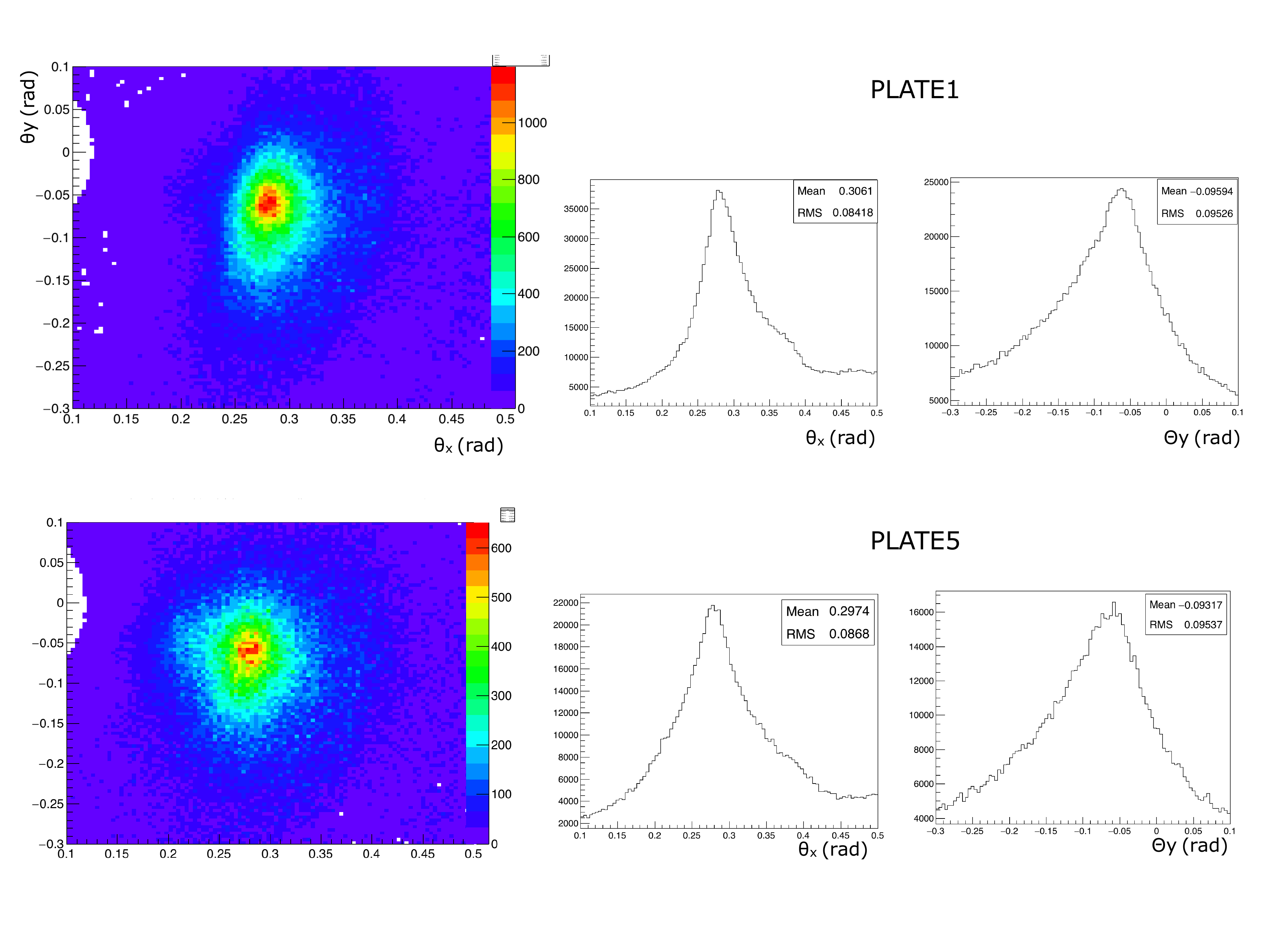}
\caption[Caption scatter plot ADC2] {Distributions of the measured inclinations of the observed particle tracks, horizontally ($\theta_{x}$) and vertically, ($\theta_{y}$) in two emulsion films, from an early layer and from a layer deep in the stack, about  the 280 mrad peak;
the region close to 0 angles has been cut off for enhancing visibiity. \label{fig:ADC_plot2}}
\end{figure}

After extraction, the emulsions were developed at CERN and analysed in the emulsion laboratory in Naples. Clear signals of particle tracks appeared, consistently reproduced in all ten layers. 
Figures
~\ref{fig:ADC_plot} and 
~\ref{fig:ADC_plot2} 
show distributions of the measured inclinations of the observed particle tracks, horizontally ($\theta_{x}$) and vertically ($\theta_{y}$) in two of the ten emulsion films, at different depths in the stack. 
The stack was aligned with the emulsions orthogonal to the beam line.
 
The track density was measured to be large, about 10$^7$ /cm$^2$. 
The analysis considered only emulsion track-lets in which hits in two sensitive layers were aligned within a $\chi^{2}$ cut off, chosen the same way as for cosmic muon tracks.
Each track-let provided two measurements: the $\theta_{x}$ and $\theta_{y}$ directions.
The emulsion package was distant 5.5 m in X , and -1.0 m below the beam axis in Y.
Each of the ten emulsions independently showed two peaks in the track angle distribution, on top of a uniform background. 
One was at about 280 mrad in $\theta_{x}$ and at about -50 mrad in $\theta_{y}$.
Tracks in this peak pointed towards the beam line and indicated a source at a distance of  about 5.5m/280mrad= 20 meters upstream along the beam line, which is confirmed by the  orthogonal measurement -1m/50mrad= 20 m.
Backgrounds depend on the machine configuration; during the emulsion exposure, the dominant contribution was expected to originate in the nearest quadrupole magnet, in agreement with the measurement. 
The other peak in the distribution of track-let angles was sharp at ~20 mrad in $\theta_{x}$ and at a few mrad below 0 in $\theta_{y}$. Since the package was at 247 m from the IP, it is expected that muon track-lets will have directions 5.5m/247m=22 mrad in X and -1m/247m= -4 mrad in Y,   which is in good agreement with the observed peak.
The size of the Ag clusters along a track in the peak is compatible with that of muons, as determined in cosmic muon mesurements.
The track population in the 22 mrad peak was measured to be 10$^5$ /cm$^2$/fb$^{-1}$, independently in each layer of the package, in good agreement with the expected flux of muons from the IP in the psudorapidity range 4$<\abs{\eta}<$5 (Section 3.3).
The track population in the 280 mrad peak was determined to be  10$^6$ /cm$^2$/fb$^{-1}$. This is consistent with the charged hadron fluence measured with the Radiation Monitors. A background of 10$^7$ /cm$^2$/fb$^{-1}$
track-lets at random angles was observed to populate the emulsions uniformly.
Although in the centre of the RR53 hall the neutron flux is measured to be  ten times smaller than in the VN location covering the same pseudorapidity range, 
it is not clear how large the 
usable area will be.
This will tend to reduce the detector acceptance, already penalised because of the small azimuthal coverage. 
Another disadvantage are localized hadronic sprays
that depend on the running configuration of LHC, as observed in our measurement. \\

\section{Very Far (VF) Location}

\subsection{Geometrical Acceptance}

From the discussion in section 2 it appears that a detector covering $\abs{\eta}>$ 6.5  will have a high flux of energetic neutrinos from b and c decays. 
In references
\cite{winter, derujula}
a detector  situated at the end of the LHC straight section within a cone with an opening angle of 2  mrad was envisaged.
A similar topology was studied also in Ref.
\cite{Park}.
This condition is not possible in the CMS intersection region, but it is possible near ATLAS. 
The cavern of tunnel TI18, used for injection of LEP beams and now in disuse, intercepts the prolongation of the beam axis  (PBA), 480 meters from the ATLAS IP, at the beginning of the collider's arc, downstream of the first bending dipoles.
The existing space is limited in length to a few meters, and the floor lies a 20-50 cm higher than the PBA. For the sake of site comparison,
a small size detector with a half-cylinder shape, only the upper half, with axis at the bottom along the PBA, with a radius of 1.2 meter, 
and a mass of 1.4$\times$10$^{27}$ nucleons/cm$^2$,  2 meters of lead, 10 ton/m,  is considered. 
The detector would cover the region of $\abs{\eta}>$ 6.7;
neutrinos have very high energies, as shown in table
~\ref{tab:fluences}, 
therefore the $\nu$N interaction cross section is large, hence the detector size can be small. 

A drawback for a  detector in this location  is that it will not be  possible to exploit the contribution from leptonic W decays. 
 
\subsection{Neutrino  Interaction Rate }

\begin{figure} [h]
\centering
\includegraphics[width=0.7\textwidth]{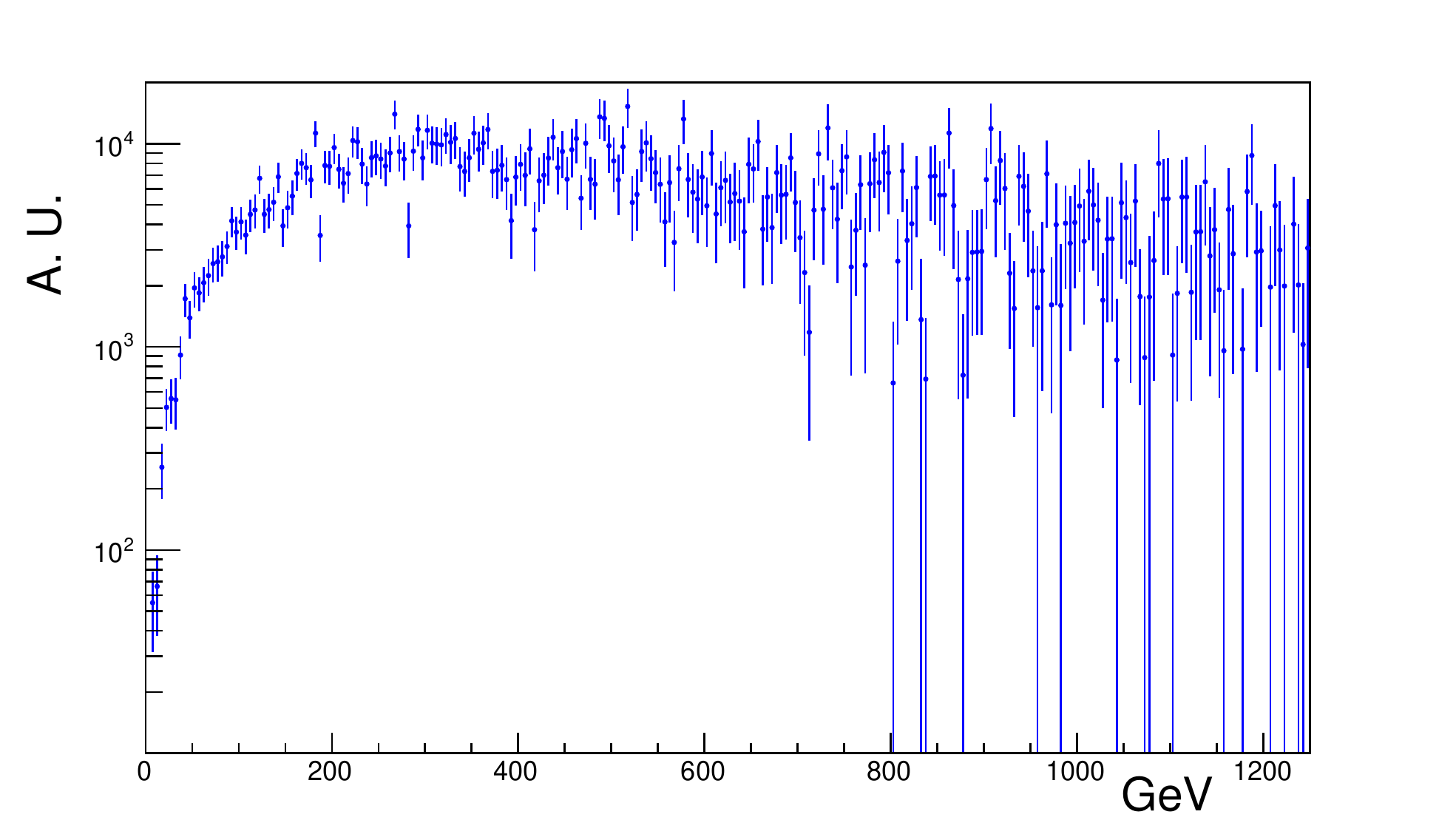}
\caption[Caption scatter plot bc] {Energy spectra for neutrinos interacting in a detector  in the VF location. \label{fig:spectraVF}}
\end{figure}

The energy spectrum of neutrinos in $\abs{\eta}>$6.7 is much harder than in the F location.  
Moreover, the small size of the detector favours neutrino interactions at higher energy, which have larger interaction cross section on nucleons. The energy spectrum of interacting neutrinos is shown
in Fig. 
~\ref{fig:spectraVF}.
Event weights are applied that take into account 
the $\nu$N interaction cross-section dependence on energy and on neutrino type
(Section 3).
Table ~\ref{tab:veryfar}  summarizes the expected neutrino interactions for 3000 fb$^{-1}$ of pp collision events, separately for heavy quark production and for W production.

\begin{table} [h]
\begin{center}
\topcaption{ Expected neutrino interactions in the very far (VF) location, for 3000 fb$^{-1}$. }
\label{tab:veryfar}
\begin{tabular} {lr} \hline 
& $\eta>$6.7 , +z end \\ \hline
Neutrinos from c and b production in pp collisions at $\sqrt{s}$=14 TeV  &  \\ \hline
total neutrino interactions & 2.8$\times$ 10$^{5}$ \\
neutrino interactions with E$>$100 GeV & 2.6$\times$ 10$^{5}$\\ 
neutrino interactions with E$>$300 GeV & 2.2$\times$ 10$^{5}$\\ \hline
$\tau$ neutrino interactions & 8700 \\ \hline \hline
Neutrinos from W production in pp collisions at $\sqrt{s}$=14 TeV  &  \\ \hline
total neutrino interactions & 2 \\
neutrino interactions with E$>$100 GeV & 1.7\\ 
neutrino interactions with E$>$300 GeV & 1.5\\ \hline 
$\tau$ neutrino interactions & 0.4\\
\end{tabular}
\end{center}
\end{table}

The contribution from leptonic W decays is negligible. The event sample consists of neutrinos from  b and c decays, with a  hard energy spectrum
($<$E$>$= 550 GeV).
With respect to the experiment targets:
\begin{itemize}
\item{\rm high energy $\nu$N cross-secion: }
a measurement of the neutrino  interaction cross section on nucleon at  energy $>$  300 GeV can be perfomed to a  1\% accuracy, and to 
a few \%  at 1 TeV. 
\item{\rm $\tau$ neutrino sample: }
many thousands of $\tau$ neutrinos can be observed, with  energies reaching into the TeV.
\end{itemize}

Therefore, a detector placed in this location, although it lies outside the kinematical range of leptonic W decays,  has a high potential for neutrino physics

\subsection{Backgrounds, Radiation Environment, and In-situ Measurement}

The backgrounds in the TI18 cavern were carefully investigated in preparation for the FASER experiment
\cite{FASER}. 
Both simulations and in-situ measurements were performed. The equipment was similar to ours: an emulsion-lead package and 
radiation monitors. The location envisaged for the FASER detector in TI18 was similar to the neutrino detector considered here, placed along the PBA and with a small transverse section.The simulations show that the particles reaching the location are dominantly muons and neutrinos. Muons come from IP but also from secondary particles  interacting along the beam line. 
The charged particle flux was measured to be about 10$^4$/cm$^2$/fb$^{-1}$. 
The radiation level was found to be low, a  neutron flux of a few $\times$10$^6$ /cm$^2$/fb$^{-1}$ was measured.
A track density in the emulsions at the 10$^4$ /cm$^2$/fb$^{-1}$ level was observed.

These values are a factor of 10, and up to 100, smaller than our  measurement in the F location.

\newpage

\section{Summary and Outlook}

\begin{figure} [h]
\centering
\includegraphics[width=1.0\textwidth]{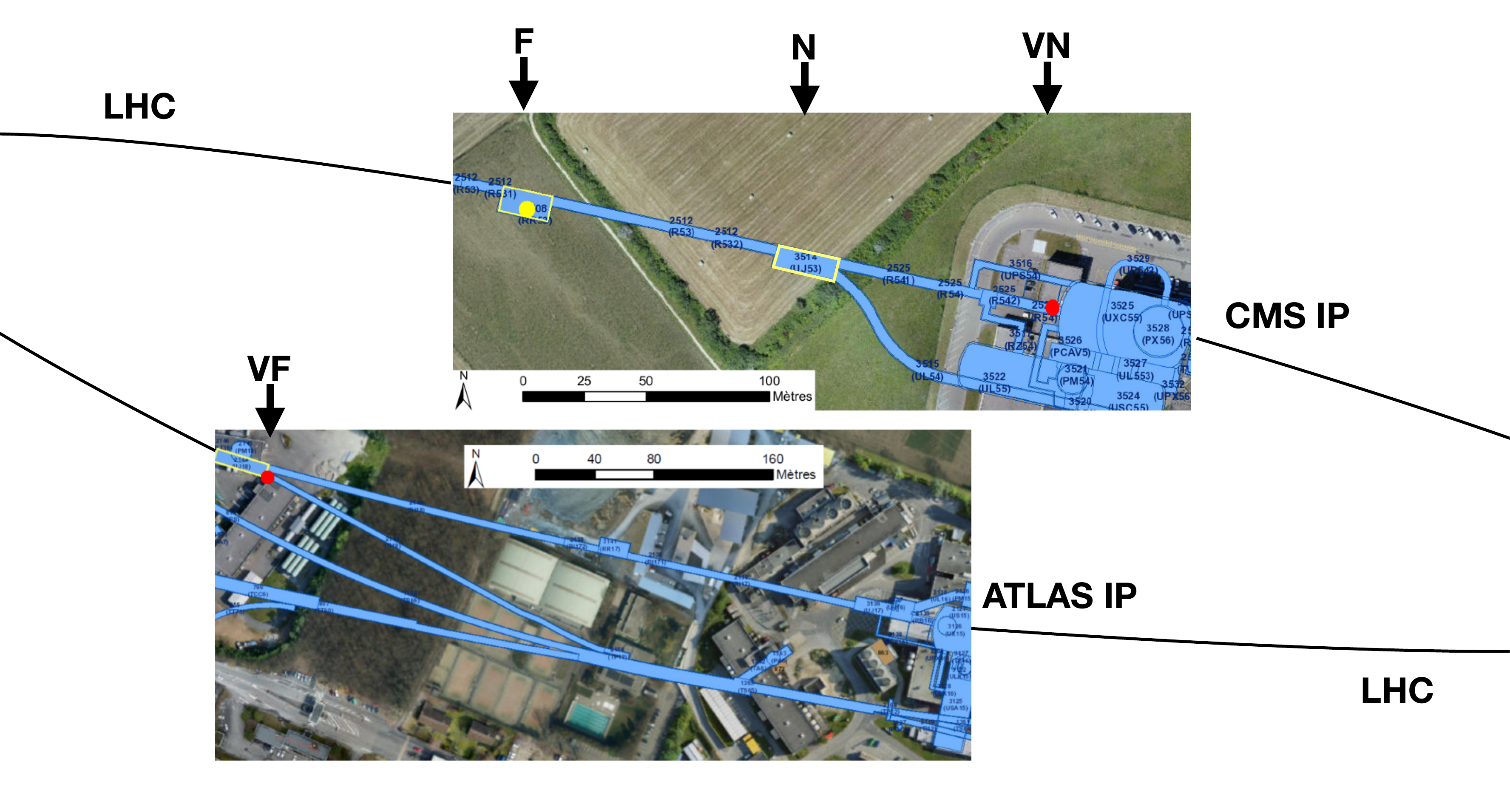}
\caption[Caption foursites] {Picture showing the  VN, N, F, and VF locations in LHC \label{fig:foursites}}
\end{figure}

Table 
~\ref{tab:bigtable}  
summarizes expected backgrounds and neutrino interactions in the four locations, recalled in Fig.
~\ref{fig:foursites}.

\begin{table} [h]
\begin{center}
\topcaption{ Expected backgrounds and  neutrino interactions in the four locations. }
\label{tab:bigtable}
\begin{tabular} {lrrrr} \hline 
& & & & \\
& VN (Q1) & N (UJ53) & F (RR53) & VF (TI18)\\ \hline
& & & & \\
muon fluence /cm$^{2}$/fb$^{-1}$ &6$\times$10$^5$   &3$\times$10$^5$  &10$^5$  &10$^2$$-$10$^4$   \\
charged hadron fluence /cm$^{2}$/fb$^{-1}$  &$>$10$^7$  &$>$10$^6$   &10$^6$$-$10$^7$   &10$^4$$-$10$^6$  \\
thermal neutron fluence /cm$^{2}$/fb$^{-1}$   &$>>$10$^8$  &$>$10$^8$   &10$^7$$-$10$^8$   &10$^6$$-$10$^7$\\ \hline
& & & & \\
L$_{int}$ = 3000 fb$^{-1}$ & & & & \\ \hline 
& & & & \\
detector volume m$^3$ &4.4  &60 &225 &11  \\ 
weight ton/m & 10  &20 & 40  & 10  \\
nucleons/cm$^2$ & 4.1$\times$10$^{27}$ &4.0$\times$10$^{27}$ & 3.0$\times$10$^{27}$ &1.4$\times$10$^{27}$ \\
& & & & \\
accuracy on $\sigma_{{\nu}N}$  & & & & \\
for neutrinos with E$>$300 GeV & 2\%  & 40\%  & 10\%  & 1\% \\ 
for neutrinos with E$\sim$ 1 TeV & 10\%  & NO  & 20\%  & 2\% \\ 
tagged leptonic W decays & 320 & 13 & 30 & 2 \\ 
& & & & \\
observed $\tau$ neutrino interactions  & 8500 & 240 & 880 &8700 \\ \hline \hline
& & & & \\
\end{tabular}
\end{center}
\end{table}

The VN and VF locations have an outstanding potential for neutrino Physics.
An experiment in VN would be sensitive to leptonic W decays where the charged lepton is observed in CMS;
however the background level makes the VN location prohibitive.
The VF location shows a sustainable level of background; in particular the thermal neutron fluence would allow use of liquid scintillator detectors,
as in reactor neutrino experiments
\cite{BUGEY}.
The flux of cosmic muons will be large with respect to deep underground neutrino experiments, but the event rate will be low because of the 
much smaller detector size.
Thus, the VF location appears to qualify for hosting a neutrino experiment. 

It should be noted that the nuclear emulsions in the F location stood 5.4  fb$^{-1}$, 
reaching a track density at the limit of the analysis capability. 
In the VF location, the background level would allow those same emulsions to stand 50-100 fb$^{-1}$ 
and to still remain within the track density limit for analysis.
This opens an interesting opportunity for the forthcoming LHC run in 2021-2023, which is expected to deliver 300 fb$^{-1}$.\\
A small size detector made of emulsions and lead, located in TI18, with 1\% of the volume as for the VF case study,  
would collect a thousand neutrino interactions
with energy ranging from 100 to 1000 GeV. 
The $\nu$N cross section at high energy would be measured  to 5\%, and to 30\% for tau neutrinos.
The lower track density required by the emulsion can be achieved by replacing the emulsions 
three to five times per year. 
This unique opportunity  that LHC offers in Run3 will be the subject of further studies, including exploration of other detector technologies. 

\begin{acknowledgments}
We would like to thank  I. Ajguirey, A. Ball, A. Benvenuti, F. Cerutti, A. Dabrowski,  D. Dattola, F. Gasparini, V. Klyukhin, M.  Komatsu. S. Mallows, A. Perrotta, I. Redondo, T. Rovelli, G. Sirri, V. Tioukov and W. Zeuner for numerous discussions and help with various parts of this study, and the CERN EN/SMM-RME groups for the support in the measurements.
\end{acknowledgments}

\bibliography{auto_generated}   

\end{document}